\documentclass[sigconf]{acmart}

\usepackage{booktabs}
\usepackage{tabularx}
\usepackage{listings}
\usepackage{xspace}
\usepackage{tikz}
\usepackage{balance}
\usetikzlibrary{arrows.meta,positioning,fit,calc}

\usepackage{booktabs}
\usepackage{tabularx}
\usepackage{array}
\usepackage{ragged2e}
\usepackage{textcomp}
\usepackage{xcolor}

\newcolumntype{Y}{>{\RaggedRight\arraybackslash}X}
\newcolumntype{P}[1]{>{\RaggedRight\arraybackslash}p{#1}}

\newcommand{\tool}{\textsc{CoCoMUT}\xspace}
\newcommand{\jsonl}{\textsc{JSONL}\xspace}
\newcommand{\spoon}{\textsc{Spoon}\xspace}
\newcommand{\sootup}{\textsc{SootUp}\xspace}

\newcommand{\ignore}[1]{}

\newcommand{\EvalReposTotal}{20}
\newcommand{\EvalReposMaven}{10}
\newcommand{\EvalReposGradle}{10}

\newcommand{\EvalMavenBuildSuccess}{10}
\newcommand{\EvalGradleBuildSuccess}{10}
\newcommand{\EvalTotalBuildSuccess}{20}
\newcommand{\EvalMavenBytecodeAvailable}{10}
\newcommand{\EvalGradleBytecodeAvailable}{10}
\newcommand{\EvalTotalBytecodeAvailable}{20}
\newcommand{\EvalMavenCallGraphAvailable}{10}
\newcommand{\EvalGradleCallGraphAvailable}{10}
\newcommand{\EvalTotalCallGraphAvailable}{20}

\newcommand{\EvalMavenMethods}{43,508}
\newcommand{\EvalGradleMethods}{13,004}
\newcommand{\EvalTotalMethods}{56,512}
\newcommand{\EvalMavenRows}{43,508}
\newcommand{\EvalGradleRows}{13,004}
\newcommand{\EvalTotalRows}{56,512}
\newcommand{\EvalMavenCallEdges}{321,806}
\newcommand{\EvalGradleCallEdges}{64,242}
\newcommand{\EvalTotalCallEdges}{386,048}

  \newcommand{\EvalMavenRuntimeMin}{9}
  \newcommand{\EvalMavenRuntimeAvg}{88}
  \newcommand{\EvalMavenRuntimeMax}{275}
  \newcommand{\EvalGradleRuntimeMin}{22}
  \newcommand{\EvalGradleRuntimeAvg}{42}
  \newcommand{\EvalGradleRuntimeMax}{68}
  \newcommand{\EvalTotalRuntimeMin}{9}
  \newcommand{\EvalTotalRuntimeAvg}{65}
  \newcommand{\EvalTotalRuntimeMax}{275}

\newcommand{\EvalMavenTargetUri}{321,806}
\newcommand{\EvalGradleTargetUri}{64,242}
\newcommand{\EvalTotalTargetUri}{386,048}
\newcommand{\EvalMavenMethodUri}{260,096}
\newcommand{\EvalGradleMethodUri}{34,146}
\newcommand{\EvalTotalMethodUri}{294,242}
\newcommand{\EvalMavenProjectAbstentions}{4,257}
\newcommand{\EvalGradleProjectAbstentions}{2,244}
\newcommand{\EvalTotalProjectAbstentions}{6,501}

\newcommand{\EvalTotalRecognizedProjectTargets}{300,743}
\newcommand{\EvalMavenNonProjectEdges}{57,453}
\newcommand{\EvalGradleNonProjectEdges}{27,852}
\newcommand{\EvalTotalNonProjectEdges}{85,305}
\newcommand{\EvalMavenProjectSourceRate}{98.4\%}
\newcommand{\EvalGradleProjectSourceRate}{93.8\%}
\newcommand{\EvalTotalProjectSourceRate}{97.8\%}

\newcommand{\EvalAuditRecords}{200}
\newcommand{\EvalAuditRepos}{10}

\newcommand{\EvalAuditPassRate}{99\%}

\title{\tool: A Tool for Code-Context Mining and Automated Dataset Generation}

\begin{document}

\author{Alessandro Botta}
\email{alessandro.botta@utdallas.edu}
% \orcid{1234-5678-9012}
% \correspondingauthor
\affiliation{%
  \institution{University of Texas at Dallas}
  \city{Richardson}
  \state{Texas}
  \country{USA}
}

\author{Shiven Garisa}
\email{garisas77@gmail.com}
\affiliation{%
  \institution{Heritage High School}
  \city{Frisco}
  \state{Texas}
  \country{USA}
}

\author{Jaya Vardhini Akurathi}
\email{jxa220064@utdallas.edu}
\affiliation{%
  \institution{University of Texas at Dallas}
  \city{Richardson}
  \state{Texas}
  \country{USA}
}

\author{Ahsanul Ameen Sabit}
\email{ahsanulameen.sabit@utdallas.edu}
\affiliation{%
  \institution{University of Texas at Dallas}
  \city{Richardson}
  \state{Texas}
  \country{USA}
}

\author{Trey Woodlief}
\email{woodlief@wm.edu}
\affiliation{%
  \institution{William \& Mary}
  \city{Williamsburg}
  \state{Virginia}
  \country{USA}
}

\author{Soneya Binta Hossain}
\email{sbhossain@utdallas.edu}
\affiliation{%
  \institution{University of Texas at Dallas}
  \city{Richardson}
  \state{Texas}
  \country{USA}
}

\begin{abstract}
Software-engineering assistants often need method-level context beyond an
isolated body, including enclosing-class information, documentation, callers,
callees, type hierarchy, and structural characteristics. Manually collecting
this context is time-consuming, inconsistent, and difficult to reproduce across
large Java projects.

We present \tool{}, a Java tool for \underline{Co}de-\underline{Co}ntext
\underline{M}ining and A\underline{ut}omated Dataset Generation. \tool{}
extracts context for a focal method or generates datasets at class, package,
or system scope. It discovers project structure, resolves build and classpath
information, constructs a SootUp static call graph, and reconciles
bytecode-level call edges with Spoon-based source extraction. Each method
record combines source, class, documentation, call-graph, and metadata
context, providing reproducible inputs for training and running learned
software-engineering techniques.

The key contribution is a reusable, task-independent pipeline that unifies
build discovery, source extraction, call-graph construction,
source--bytecode reconciliation, and versioned JSON dataset generation. The
resulting records can be consumed individually as context for a focal method or
collectively as datasets for documentation, explanation, testing, review,
repair, search, and program-comprehension workflows. We evaluate \tool{} on
\EvalReposTotal{} real-world Java repositories evenly split between Maven and
Gradle. \tool{} processed all \EvalReposTotal{} repositories, emitting
\EvalTotalRows{} method-context records and \EvalTotalCallEdges{} serialized
call edges. Among call edges whose bytecode targets belonged to project source,
\tool{} reconciled \EvalTotalProjectSourceRate{} to source method identities.
In a manual audit of \EvalAuditRecords{} randomly sampled methods across
\EvalAuditRepos{} systems, \EvalAuditPassRate{} of generated context records passed all applicable correctness checks.\\

\noindent\textbf{Demo:} \url{https://youtu.be/RCUzkCQjG30} \\
\textbf{Artifact:} \url{https://github.com/assert-lab/CoCoMUT}
\end{abstract}



\maketitle

\section{Introduction}

Many software-engineering tasks require context beyond an isolated method
body. File-level context improves code summarization
accuracy~\cite{haque2020filecontext}, while test generation and program repair
use information about the focal method, enclosing type, and related program
elements~\cite{tufano2022methods2test,li2020dlfix,hossain2025togll,hossain2025doc2oracll,hossain2024deep}.
At scale, this context is difficult to construct because it is spread across
source code, in-source documentation, type hierarchies, dependencies, and
compiled bytecode. Researchers therefore build task-specific extraction
pipelines with different assumptions about method identity, dependency
resolution, and context boundaries~\cite{hossain2025togll,hossain2025doc2oracll,schaefer2024empirical,chen2024chatunitest,wang2024hits,zhang2023repocoder,zhang2024autocoderover},
hindering comparison and reproducibility~\cite{liu2022reproducibility}.

Existing source-analysis and call-graph frameworks provide useful building
blocks, but reusable method-context extraction still requires resolving
project-specific builds and classpaths, combining source- and bytecode-level
analyses, reconciling program identities, and defining a consistent output
schema. Java adds further complications: calls to overridable methods,
including interface methods, can have multiple runtime targets, forcing static
analyses to balance precision and scalability~\cite{sundaresan2000virtual};
overloading, nested and anonymous types, inheritance, and generics complicate
stable method identification.

We present \tool{}, a Java tool for automated code-context mining and dataset
generation shown in Figure~\ref{fig:pipeline}. Given a project and selection scope, \tool{} discovers project
structure, resolves build and classpath information, builds a Spoon source
model~\cite{spoon}, constructs a SootUp static call graph~\cite{sootup}, and
reconciles bytecode call targets with source-level method identities. It emits
one versioned \jsonl{} record per selected method~\cite{jsonlines}, combining
method identity, implementation, enclosing-type context, documentation,
callers and callees, hierarchy information, structural metrics, provenance,
and metadata. Records can be inspected individually or processed as
datasets over methods, classes, packages, or entire systems.

We evaluate \tool{} on \EvalReposTotal{} real-world Java repositories, split
evenly between Maven and Gradle projects. We measure extraction robustness,
source--bytecode reconciliation for project call edges, and output quality
under manual audit.

In summary, this paper contributes:

\begin{itemize}
    \item \textbf{A task-independent method-context schema} that unifies
    stable method identity, implementation, local class context, field
    accesses, overload and sibling-method context, structured and linked
    Javadoc, type hierarchy, caller/callee context, structural metrics, and
    provenance/confidence metadata. See Table~\ref{tab:record-context}.

    \item \textbf{An automated Java context-mining pipeline} integrating
    project discovery, build/classpath resolution, Spoon source analysis,
    SootUp call-graph construction, source--bytecode reconciliation,
    configurable method selection, and versioned \jsonl{} generation.

    \item \textbf{An empirical evaluation} on \EvalReposTotal{} Java
    repositories, measuring extraction robustness, emitted records, call-graph
    availability, project-source reconciliation, and abstention categories.

    \item \textbf{A replication package} with the tool, documentation,
    evaluation artifacts, and reproduction information.
\end{itemize}

\section{\tool{} Workflow}

Figure~\ref{fig:pipeline} summarizes the \tool{} workflow. \tool{}
combines two views of the same project: source declarations modeled by
\spoon{} and bytecode methods modeled by \sootup{}. It reconciles these views
through stable method identities and emits one method-context record per
selected method. The workflow has four phases; phase~2 builds the source and
bytecode views.

\begin{description}
    \item[\textbf{1. Input and extraction request.}]
    The user provides a Java project and an extraction request, such as a
    source set, package, type, method filter, or exact focal method.

    \item[\textbf{2a. CoCoMUT source model.}]
    \tool{} uses \spoon{} to build a source model and projects it into a
    CoCoMUT source index containing stable method URIs, Javadoc, annotations,
    hierarchy information, and other source context.

    \item[\textbf{2b. CoCoMUT bytecode analysis.}]
    \tool{} analyzes compiled project bytecode and dependency artifacts with
    \sootup{}, constructing a CHA or RTA call graph with bytecode-level caller
    and callee edges.

    \item[\textbf{3. Source--bytecode reconciliation.}]
    \tool{} reconciles \spoon{} source-method keys with \sootup{} bytecode
    signatures. Unique matches receive a source \texttt{method\_uri},
    ambiguous matches retain candidate URIs, and unmatched bytecode targets
    remain visible through \texttt{target\_uri}.

    \item[\textbf{4. Versioned outputs and diagnostics.}]
    \tool{} serializes method-context records as versioned \jsonl{} and emits
    an extraction report, failure artifacts, and summary statistics.
\end{description}

% Faithful four-stage reproduction of the revised CoCoMUT pipeline.
% Derived from the original TikZ pipeline source.
%
% Requires in the preamble:
%   \usepackage{graphicx}
%   \usepackage{tikz}
%   \usetikzlibrary{arrows.meta,calc,positioning}
%
% The figure is drawn on a large natural canvas and scaled to the full text
% width.  This keeps the typography and spacing stable in ACM two-column mode.

\definecolor{CoBlue}{HTML}{1456C5}
\definecolor{CoBlueFill}{HTML}{F4F7FF}
\definecolor{CoGreen}{HTML}{138A2E}
\definecolor{CoGreenFill}{HTML}{F7FCF7}
\definecolor{CoPurple}{HTML}{6B2BA6}
\definecolor{CoPurpleFill}{HTML}{FBF8FD}
\definecolor{CoGold}{HTML}{D47A00}
\definecolor{CoGoldFill}{HTML}{FFFBF5}
\definecolor{CoOrange}{HTML}{E14F09}
\definecolor{CoOrangeFill}{HTML}{FFF9F5}
\definecolor{CoRed}{HTML}{D62620}
\definecolor{CoInk}{HTML}{202020}
\definecolor{CoGray}{HTML}{555555}

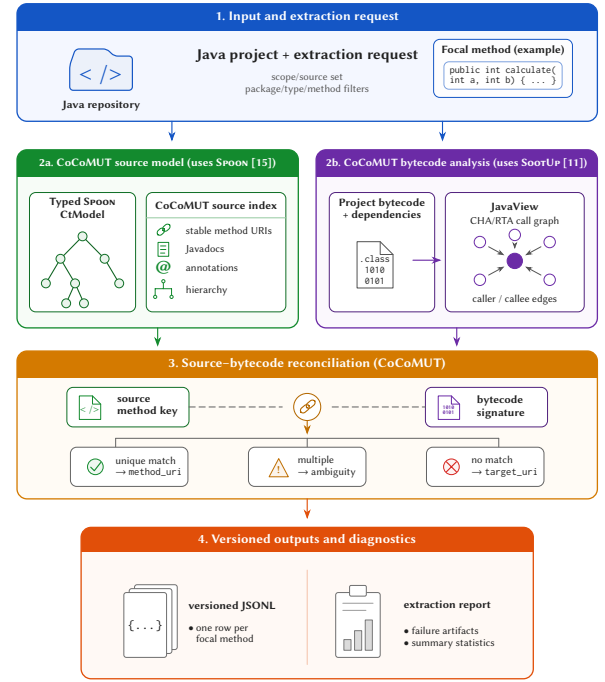
\begin{figure}[t]
\centering
\resizebox{0.45\textwidth}{!}{%
\begin{tikzpicture}[
  x=1cm,
  y=1cm,
  line cap=round,
  line join=round,
  font=\sffamily,
  outer/.style={line width=1.05pt, rounded corners=8pt},
  inner/.style={line width=0.78pt, rounded corners=5pt, fill=white},
  header/.style={
    text=white,
    font=\sffamily\bfseries\fontsize{10.5}{11.8}\selectfont,
    align=center
  },
  title/.style={font=\sffamily\bfseries\fontsize{12.0}{13.2}\selectfont, text=CoInk},
  subtitle/.style={font=\sffamily\bfseries\fontsize{9.0}{10.0}\selectfont, text=CoInk},
  body/.style={font=\sffamily\fontsize{8.5}{9.5}\selectfont, text=CoInk},
  small/.style={font=\sffamily\fontsize{7.7}{8.6}\selectfont, text=CoInk},
  mono/.style={font=\ttfamily\fontsize{8.1}{9.0}\selectfont, text=CoInk},
  flow/.style={-{Stealth[length=2.8mm,width=2.0mm]}, line width=0.90pt},
  thinflow/.style={-{Stealth[length=2.0mm,width=1.4mm]}, line width=0.70pt}
]

% Natural canvas, matching the portrait reference image.
\path[use as bounding box] (0,0) rectangle (17.0,21.0);

% Helper: each header is clipped by the rounded outer panel so all top corners
% and outline proportions remain uniform.

% ============================================================================
% 1. INPUT AND EXTRACTION REQUEST
% ============================================================================
\begin{scope}
  \clip[rounded corners=8pt] (0.35,17.35) rectangle (16.65,20.65);
  \fill[CoBlueFill] (0.35,17.35) rectangle (16.65,20.65);
  \fill[CoBlue] (0.35,19.88) rectangle (16.65,20.65);
\end{scope}
\draw[outer, draw=CoBlue] (0.35,17.35) rectangle (16.65,20.65);
\node[header] at (8.50,20.27) {1. Input and extraction request};

\node[title] at (8.50,19.18) {Java project + extraction request};
\node[body, align=center, text=CoGray] at (8.50,18.40)
  {scope/source set\\ package/type/method filters};

% Repository folder icon.
\begin{scope}[shift={(2.65,18.80)}]
  \draw[draw=CoBlue, fill=CoBlue!8, line width=0.95pt, rounded corners=2.8pt]
    (-0.82,0.12)--(-0.66,0.58)--(0.22,0.58)--(0.43,0.38)
    --(0.94,0.38)--(0.94,-0.58)--(-0.82,-0.58)--cycle;
  \node[text=CoBlue, font=\sffamily\bfseries\fontsize{17}{17}\selectfont]
    at (0.05,-0.13) {$</>$};
\end{scope}
\node[subtitle] at (2.72,17.80) {Java repository};

% Focal-method card.
\draw[inner, draw=CoBlue, fill=CoBlue!1] (12.05,18.02) rectangle (15.95,19.65);
\node[subtitle] at (14.00,19.34) {Focal method (example)};
\draw[inner, draw=CoBlue!55, fill=white, line width=0.65pt, rounded corners=4pt]
  (12.34,18.25) rectangle (15.66,19.02);
\node[mono, align=center] at (14.00,18.635)
  {public int calculate(\\[-0.1ex]  int a, int b) \{ ... \}};

% Input arrows.
\draw[flow, draw=CoBlue] (4.70,17.35)--(4.70,16.78);
\draw[flow, draw=CoBlue] (12.30,17.35)--(12.30,16.78);

% ============================================================================
% 2a. SOURCE MODEL
% ============================================================================
\begin{scope}
  \clip[rounded corners=8pt] (0.35,11.55) rectangle (8.25,16.55);
  \fill[CoGreenFill] (0.35,11.55) rectangle (8.25,16.55);
  \fill[CoGreen] (0.35,15.80) rectangle (8.25,16.55);
\end{scope}
\draw[outer, draw=CoGreen] (0.35,11.55) rectangle (8.25,16.55);
\node[font=\sffamily\bfseries\fontsize{9.15}{10.0}\selectfont, text=white, align=center] at (4.30,16.17) {2a. CoCoMUT source model (uses \spoon~\cite{spoon})};

% CtModel card.
\draw[inner, draw=CoGreen!80!black] (0.70,12.00) rectangle (3.68,15.35);
\node[subtitle, align=center] at (2.19,14.93) {Typed \spoon\\CtModel};

% Tree icon, centered with comfortable margins.
\coordinate (s0) at (2.19,14.12);
\coordinate (s1) at (1.57,13.49);
\coordinate (s2) at (2.81,13.49);
\coordinate (s3) at (1.16,12.80);
\coordinate (s4) at (1.98,12.80);
\coordinate (s5) at (3.22,12.80);
\coordinate (s6) at (1.73,12.24);
\coordinate (s7) at (2.23,12.24);
\draw[draw=CoGreen!72!black, line width=0.86pt]
  (s0)--(s1) (s0)--(s2) (s1)--(s3) (s1)--(s4)
  (s2)--(s5) (s4)--(s6) (s4)--(s7);
\foreach \p in {s0,s1,s2,s3,s4,s5,s6,s7}
  \filldraw[fill=CoGreen!9, draw=CoGreen!80!black, line width=0.78pt] (\p) circle (0.125);

% Source-index card.
\draw[inner, draw=CoGreen!80!black] (3.98,12.00) rectangle (7.90,15.35);
\node[subtitle] at (5.94,15.00) {CoCoMUT source index};
\draw[draw=CoGreen!45, line width=0.62pt] (3.98,14.64)--(7.90,14.64);

% Source-index rows.
\begin{scope}[shift={(4.43,14.28)}]
  % link
  \begin{scope}[rotate=42]
    \draw[draw=CoGreen, line width=0.82pt, rounded corners=2.5pt]
      (-0.17,-0.07) rectangle (0.07,0.08);
    \draw[draw=CoGreen, line width=0.82pt, rounded corners=2.5pt]
      (0.02,-0.09) rectangle (0.26,0.06);
  \end{scope}
  \node[body, anchor=west] at (0.55,0) {stable method URIs};

  % document
  \draw[draw=CoGreen, fill=white, line width=0.75pt] (-0.12,-0.72) rectangle (0.18,-0.33);
  \draw[draw=CoGreen, line width=0.5pt]
    (-0.05,-0.43)--(0.10,-0.43) (-0.05,-0.53)--(0.10,-0.53) (-0.05,-0.63)--(0.07,-0.63);
  \node[body, anchor=west] at (0.55,-0.52) {Javadocs};

  % @ annotation
  \node[text=CoGreen, font=\sffamily\bfseries\fontsize{13}{13}\selectfont] at (0.03,-1.04) {@};
  \node[body, anchor=west] at (0.55,-1.04) {annotations};

  % hierarchy
  \draw[draw=CoGreen, line width=0.75pt]
    (0.03,-1.45)--(0.03,-1.63) (-0.20,-1.63)--(0.26,-1.63)
    (-0.20,-1.63)--(-0.20,-1.82) (0.26,-1.63)--(0.26,-1.82);
  \filldraw[fill=white, draw=CoGreen, line width=0.65pt] (0.03,-1.42) circle (0.045);
  \filldraw[fill=white, draw=CoGreen, line width=0.65pt] (-0.20,-1.86) circle (0.045);
  \filldraw[fill=white, draw=CoGreen, line width=0.65pt] (0.26,-1.86) circle (0.045);
  \node[body, anchor=west] at (0.55,-1.67) {hierarchy};
\end{scope}

\draw[flow, draw=CoGreen] (4.30,11.55)--(4.30,11.02);

% ============================================================================
% 2b. BYTECODE ANALYSIS
% ============================================================================
\begin{scope}
  \clip[rounded corners=8pt] (8.75,11.55) rectangle (16.65,16.55);
  \fill[CoPurpleFill] (8.75,11.55) rectangle (16.65,16.55);
  \fill[CoPurple] (8.75,15.80) rectangle (16.65,16.55);
\end{scope}
\draw[outer, draw=CoPurple] (8.75,11.55) rectangle (16.65,16.55);
\node[font=\sffamily\bfseries\fontsize{8.95}{9.8}\selectfont, text=white, align=center] at (12.70,16.17) {2b. CoCoMUT bytecode analysis (uses \sootup~\cite{sootup})};

% Project bytecode card.
\draw[inner, draw=CoPurple!75!black] (9.12,12.00) rectangle (12.08,15.35);
\node[subtitle, align=center] at (10.60,14.88) {Project bytecode\\+ dependencies};

\draw[draw=CoGray, fill=white, line width=0.82pt]
  (9.90,12.67)--(9.90,13.97)--(10.55,13.97)--(10.91,13.61)
  --(10.91,12.67)--cycle;
\draw[draw=CoGray, line width=0.62pt] (10.55,13.97)--(10.55,13.61)--(10.91,13.61);
\node[mono, align=center] at (10.40,13.18) {.class\\1010\\0101};

% JavaView card.
\draw[inner, draw=CoPurple!75!black] (12.38,12.00) rectangle (16.28,15.35);
\node[subtitle] at (14.33,14.92) {JavaView};
\node[body] at (14.33,14.52) {CHA/RTA call graph};

\node[circle, draw=CoPurple, fill=CoPurple, line width=0.82pt, minimum size=0.43cm, inner sep=0pt]
  (mid) at (14.33,13.43) {};
\foreach \name/\xx/\yy in {
  c1/13.35/13.96,c2/14.33/14.16,c3/15.31/13.96,
  c4/13.35/12.90,c5/15.31/12.90}{
  \node[circle, draw=CoPurple, fill=white, line width=0.75pt, minimum size=0.32cm, inner sep=0pt]
    (\name) at (\xx,\yy) {};
}
\foreach \n in {c1,c2,c3,c4,c5}
  \draw[thinflow, draw=CoGray, shorten <=3pt, shorten >=3pt] (\n)--(mid);
\node[body] at (14.33,12.34) {caller / callee edges};

\draw[flow, draw=CoPurple] (11.47,13.45)--(12.27,13.45);
\draw[flow, draw=CoPurple] (12.70,11.55)--(12.70,11.02);

% ============================================================================
% 3. SOURCE-BYTECODE RECONCILIATION
% ============================================================================
\begin{scope}
  \clip[rounded corners=8pt] (0.35,6.75) rectangle (16.65,10.90);
  \fill[CoGoldFill] (0.35,6.75) rectangle (16.65,10.90);
  \fill[CoGold] (0.35,10.17) rectangle (16.65,10.90);
\end{scope}
\draw[outer, draw=CoGold] (0.35,6.75) rectangle (16.65,10.90);
\node[header] at (8.50,10.535) {3. Source--bytecode reconciliation (CoCoMUT)};

% Source method key.
\draw[inner, draw=CoGreen!80!black] (1.75,8.77) rectangle (5.18,9.88);
\draw[draw=CoGreen, fill=white, line width=0.76pt]
  (2.10,8.98)--(2.10,9.64)--(2.47,9.64)--(2.70,9.41)--(2.70,8.98)--cycle;
\draw[draw=CoGreen, line width=0.56pt] (2.47,9.64)--(2.47,9.41)--(2.70,9.41);
\node[text=CoGreen, font=\sffamily\bfseries\fontsize{7}{11}\selectfont] at (2.39,9.25) {$</>$};
\node[subtitle, align=left, anchor=west] at (3.05,9.33) {source\\ method key};

% Bytecode signature, simple document/binary symbol.
\draw[inner, draw=CoPurple!75!black] (11.82,8.77) rectangle (15.25,9.88);
\draw[draw=CoPurple, fill=white, line width=0.76pt]
  (12.18,8.98)--(12.18,9.64)--(12.55,9.64)--(12.78,9.41)--(12.78,8.98)--cycle;
\draw[draw=CoPurple, line width=0.56pt] (12.55,9.64)--(12.55,9.41)--(12.78,9.41);
\node[font=\ttfamily\bfseries\fontsize{4.5}{4.5}\selectfont, text=CoPurple, align=center]
  at (12.47,9.25) {1010\\0101};
\node[subtitle, align=left, anchor=west] at (13.15,9.33) {bytecode\\ signature};

% Dashed join and central link.
\draw[draw=CoInk!72, dash pattern=on 5pt off 3pt, line width=0.78pt]
  (5.18,9.33)--(7.80,9.33);
\draw[draw=CoInk!72, dash pattern=on 5pt off 3pt, line width=0.78pt]
  (9.20,9.33)--(11.82,9.33);
\draw[draw=CoGold!85!black, fill=white, line width=0.82pt] (8.50,9.33) circle (0.38);
\begin{scope}[shift={(8.50,9.33)}, rotate=40]
  \draw[draw=CoGold!85!black, line width=0.88pt, rounded corners=2.6pt]
    (-0.23,-0.075) rectangle (0.06,0.08);
  \draw[draw=CoGold!85!black, line width=0.88pt, rounded corners=2.6pt]
    (-0.02,-0.08) rectangle (0.27,0.075);
\end{scope}

% Branch to outcomes.
\draw[flow, draw=CoGold!90!black] (8.50,8.93)--(8.50,8.55);
\draw[draw=CoInk!70, line width=0.72pt] (3.12,8.45)--(13.88,8.45);
\draw[draw=CoInk!70, line width=0.72pt] (3.12,8.45)--(3.12,8.20);
\draw[draw=CoInk!70, line width=0.72pt] (8.50,8.45)--(8.50,8.20);
\draw[draw=CoInk!70, line width=0.72pt] (13.88,8.45)--(13.88,8.20);

% Outcome cards.
\draw[inner, draw=CoInk!70] (1.85,7.10) rectangle (5.15,8.20);
\draw[draw=CoGreen, fill=CoGreen!7, line width=0.80pt] (2.55,7.65) circle (0.22);
\draw[draw=CoGreen, line width=0.90pt] (2.42,7.64)--(2.52,7.54)--(2.70,7.77);
\node[body, align=left, anchor=west] at (3.00,7.66)
  {unique match\\[-0.10ex]$\rightarrow$ \texttt{method\_uri}};

\draw[inner, draw=CoInk!70] (6.85,7.10) rectangle (10.15,8.20);
\draw[draw=CoGold!90!black, fill=CoGold!8, line width=0.80pt]
  (7.43,7.37)--(7.70,7.88)--(7.97,7.37)--cycle;
\node[font=\sffamily\bfseries\fontsize{8.5}{8.5}\selectfont, text=CoGold!90!black]
  at (7.70,7.57) {!};
\node[body, align=left, anchor=west] at (8.12,7.66)
  {multiple\\[-0.10ex]$\rightarrow$ ambiguity};

\draw[inner, draw=CoInk!70] (11.85,7.10) rectangle (15.15,8.20);
\draw[draw=CoRed, fill=CoRed!6, line width=0.80pt] (12.55,7.65) circle (0.22);
\draw[draw=CoRed, line width=0.90pt] (12.41,7.51)--(12.69,7.79) (12.41,7.79)--(12.69,7.51);
\node[body, align=left, anchor=west] at (13.00,7.66)
  {no match\\[-0.10ex]$\rightarrow$ \texttt{target\_uri}};

\draw[flow, draw=CoOrange] (8.50,6.75)--(8.50,6.18);

% ============================================================================
% 4. VERSIONED OUTPUTS AND DIAGNOSTICS (COMPRESSED)
% ============================================================================
\begin{scope}
  % 1. Changed bottom coordinate from 1.10 to 1.70 to make the rectangle less tall
  \clip[rounded corners=8pt] (2.15,1.70) rectangle (14.85,5.95);
  \fill[CoOrangeFill] (2.15,1.70) rectangle (14.85,5.95);
  \fill[CoOrange] (2.15,5.25) rectangle (14.85,5.95);
\end{scope}
% Changed bottom coordinate from 1.10 to 1.70
\draw[outer, draw=CoOrange] (2.15,1.70) rectangle (14.85,5.95);
\node[header] at (8.50,5.60) {4. Versioned outputs and diagnostics};

% Divider: Changed bottom starting Y from 1.60 to 2.10 to match new box floor
\draw[draw=CoOrange!35, line width=0.62pt] (8.50,2.10)--(8.50,4.75);

% Versioned JSONL stack: Shifted base Y up from 2.00 to 2.30
\foreach \dx/\dy in {0.30/0.22,0.15/0.11,0/0} {
  \draw[draw=CoInk!65, fill=white, line width=0.76pt, rounded corners=2pt]
    (3.35+\dx,2.30+\dy) rectangle (4.55+\dx,4.12+\dy);
}
\node[font=\ttfamily\fontsize{11}{11}\selectfont, text=CoInk] at (3.95,3.18) {\{...\}}; % Y changed from 2.88 to 3.18
\node[subtitle, anchor=west] at (5.05,3.75) {versioned JSONL};                       % Y changed from 3.55 to 3.75
\node[body, align=left, anchor=west] at (5.05,2.95)                                  % Y changed from 2.65 to 2.95
  {$\bullet$ one row per\\[-0.08ex]\hspace{0.22cm}focal method};

% Extraction-report clipboard: Shifted base Y up from 2.00 to 2.30
\draw[draw=CoInk!65, fill=white, line width=0.76pt, rounded corners=2pt]
  (9.30,2.30) rectangle (10.55,4.12);
\draw[draw=CoInk!65, fill=white, line width=0.62pt, rounded corners=1.2pt]
  (9.69,4.02) rectangle (10.16,4.32);

% Spacing/Collision Fix: Adjusted inner horizontal lines
\draw[draw=CoInk!45, line width=0.50pt] (9.50,3.82)--(10.35,3.82) (9.50,3.62)--(10.15,3.62);

% Spacing/Collision Fix: Scaled and shifted bar heights safely away from lines
\draw[draw=CoInk!60, fill=CoInk!16, line width=0.50pt] (9.53,2.50) rectangle (9.73,2.80);
\draw[draw=CoInk!60, fill=CoInk!16, line width=0.50pt] (9.83,2.50) rectangle (10.03,3.05);
\draw[draw=CoInk!60, fill=CoInk!16, line width=0.50pt] (10.13,2.50) rectangle (10.33,3.35);

% Text labels: Shifted Y positions up by 0.20 to align with the new icon centers
\node[subtitle, anchor=west] at (11.10,3.75) {extraction report};                    % Y changed from 3.55 to 3.75
\node[body, align=left, anchor=west] at (11.10,2.85)                                 % Y changed from 2.65 to 2.85
  {$\bullet$ failure artifacts\\[-0.08ex]$\bullet$ summary statistics};
\end{tikzpicture}%
}
\caption{\tool{} execution on a Java system.}
\Description{Four-stage CoCoMUT workflow. A Java project and extraction request feed a \spoon-based source model and a \sootup bytecode analysis. CoCoMUT reconciles a source method key with a bytecode signature, distinguishes unique, ambiguous, and unmatched targets, and emits versioned JSONL records together with extraction diagnostics.}
\label{fig:pipeline}
\end{figure}

\subsection{Project Analysis and Execution Modes}

\tool{} first identifies the project layout and build metadata needed for
source parsing and bytecode analysis. For Maven and Gradle projects, \tool{}
can build fresh bytecode or consume project artifacts already present in
conventional layouts. The pipeline requires compiled bytecode for the source
project; partial bytecode, e.g., from compilation failures, is insufficient.
We leave partial-bytecode extraction to future work.

\tool{} does not run project tests or application entry points directly.
However, Maven and Gradle builds may execute repository-controlled plugins and
scripts, so untrusted repositories should be analyzed in an isolated
environment.

\subsection{Method Selection and Stable Identity}

After project analysis, \tool{} opens one source-analysis session and
enumerates source methods and constructors. It maintains two sets: the
\emph{analysis universe}, containing all discovered source methods for identity
resolution, and the \emph{focal set}, containing only methods selected for
output. Each source method receives a canonical URI of the form: %This separation ensures that narrowing the output scope does not change whether edges to other project methods can be recognized. 

\begin{center}
\small
\path{src/main/java/p/Foo.java#p.Foo.parse(java.lang.String):int}
\end{center}

The URI combines repository-relative source location, qualified declaring type,
method or constructor name, erased parameter types, and erased return type. It
distinguishes overloads, nested types, generic erasures, and constructors, and
is also used for exact method selection. Type and package targets use the same
\texttt{path\#symbol} convention.

\subsection{Source-Level Context Extraction}

\tool{} uses \spoon{} as its source-code front end~\cite{spoon}. \spoon{}
parses Java source into a typed \texttt{CtModel} containing program elements
such as types, methods, constructors, fields, references, statements, and
comments.

\tool{} builds this model once per request and projects it into
framework-independent \texttt{SourceMethod} and \texttt{SourceContext} records.
A request-local index maps stable CoCoMUT URIs to source records and their
corresponding \spoon{} \texttt{CtExecutable}s, qualified type names to
\spoon{} types, and declaring types to their methods and fields. This
projection establishes stable source identity by normalizing source-level and
erased Java types, distinguishing overloads and constructors, handling nested
declarations, preserving source positions, and classifying methods by source
set.

Documentation parsing follows the standard Javadoc syntax defined by the
Oracle/JDK documentation-comment specification~\cite{javadoc17}. \tool{} uses
\spoon{}'s \texttt{spoon-javadoc} module for block and inline tags. For
program-element references such as \texttt{@see} and \texttt{\{@link ...\}},
\tool{} maps typed \spoon{} references to CoCoMUT method, field, or type URIs
when the target belongs to the analyzed project.

For \texttt{\{@inheritDoc\}}, \tool{} does not silently merge inherited text
into the child method's structured tags. Instead, it reports whether inherited
documentation candidates exist and exposes those candidates explicitly.

\subsection{Call-Graph Construction and Source--Bytecode Reconciliation}

\tool{} builds a static call graph from compiled project bytecode using
\sootup{}. Users can select rapid type analysis (RTA), the default, or
class-hierarchy analysis (CHA).

\sootup{} and \spoon{} identify methods at different levels: \sootup{} reports
bytecode signatures, while \spoon{} reports source declarations. These
identities can differ because of generic type erasure, synthetic methods, and
other source-to-bytecode transformations. Therefore, \tool{} links a bytecode
target to a source method only when the match is unique.

For each caller/callee relation reported by \sootup{}, \tool{} records the
  related method's bytecode identity as \texttt{target\_uri}.
 and then attempts to add a source-level
\texttt{method\_uri}. Multiple compatible source declarations produce an
explicit ambiguity with candidate URIs; no compatible source declaration
leaves \texttt{method\_uri} empty.

This separation prevents false source links. External targets and
compiler-generated targets, such as synthetic bridge methods or lambda
artifacts, keep their bytecode identity, while \texttt{method\_uri} appears
only for project-source targets that \tool{} resolves deterministically.

\subsection{Output and Metadata}

\tool{} writes one deterministic \jsonl{} record for each focal method.
Table~\ref{tab:record-context} summarizes the five context families in each
record: identity/source, local class, documentation, call-graph, and
metadata. Full schema details are available at
\url{https://github.com/assert-lab/CoCoMUT/tree/main/schemas}. Output filenames
include a request fingerprint, so different extraction configurations produce
distinct artifacts. By default, \tool{} writes artifacts outside the analyzed
repository, avoiding target-project pollution and simplifying repeated
experiments.

\begin{table}[t]
\caption{Context families emitted in a \tool{} method record; full schema can be found at~\cite{cocomut}}
\label{tab:record-context}
\small
\begin{tabular}{@{}p{0.27\linewidth}p{0.65\linewidth}@{}}
\toprule
Family & Examples \\
\midrule
Identity and source &
Stable URI, signature, erased types, source set, position, parameters,
annotations, throws, code, LOC, cyclomatic complexity. \\
Local class context &
Class Javadoc, hierarchy, sibling methods, overload group, field reads and
writes. \\
Documentation context &
Structured Javadoc, \texttt{@param}, \texttt{@return}, \texttt{@throws},
\texttt{@since}, deprecation text, inline links, \texttt{@see},
inherited-documentation candidates. \\
Call context &
Caller/callee edges, bytecode \texttt{target\_uri}, source
\texttt{method\_uri}, raw signature, target kind, nested project-method
context when available. \\
Provenance and confidence &
Schema version, selection scope, source backend, call-graph algorithm,
resolution status, unresolved reason, context confidence. \\
\bottomrule
\end{tabular}
\end{table}

\subsection{Implementation}

\tool{} is implemented as a Maven multi-module Java project and exposes the
pipeline through a CLI, a shaded executable JAR, and a Java service API. It
builds on \spoon{} and \texttt{spoon-\allowbreak javadoc} for typed source and
documentation models, \sootup{} for bytecode loading and call graphs, and
Jackson for serializing versioned JSON records~\cite{spoon,sootup,jackson_repo}.

The evaluated release runs on JDK~17 or newer. Source parsing is configured
for Java source levels up to Java~25, while call-graph construction is limited
by the bytecode versions supported by \sootup{}, up to Java~21 in the evaluated
release. %Projects outside these bounds are reported through failure and provenance metadata rather than silently downgraded to source-only results.

\section{Evaluation}
\label{sec:evaluation}
We evaluate \tool{} as a context extraction tool on \EvalReposTotal{} real-world Java
repositories: \EvalReposMaven{} Maven and \EvalReposGradle{} Gradle. Experiments
ran on a Linux x86-64 consumer machine with OpenJDK~17.0.19, Maven~3.9.16, and
Gradle~9.5.1. Each repository was checked out once and analyzed with
main-source, all-method scope, RTA, and build execution enabled. The
reproduction script records the environment, parses extraction reports,
validates method-context rows, and counts serialized caller/callee entries. \\

We ran the following command for all systems.

\begin{lstlisting}[basicstyle=\ttfamily\scriptsize,aboveskip=2pt,belowskip=2pt]
./bin/cocomut --project <project> --scope all \
--source-set main --call-graph rta \
--allow-build --output-dir <output>
\end{lstlisting}

The evaluation addresses three research questions:

\begin{itemize}
  \item[\textbf{RQ1.}] \textbf{Build-ecosystem robustness.}
  Can \tool{} produce method-context records across Maven and Gradle projects?

  \item[\textbf{RQ2.}] \textbf{Source--bytecode reconciliation.}
  How often can \tool{} deterministically link bytecode call targets to
  source-level project methods, and when does it abstain?

  \item[\textbf{RQ3.}] \textbf{Output quality.}
  Do manually audited records contain the intended method identity,
  documentation references, caller/callee context, and inherited-documentation
  metadata?
\end{itemize}

  \subsection{RQ1: Build-Ecosystem Robustness}

  \begin{table}[h]
  \centering
 \caption{RQ1 results. B/BC/CG reports successful build, project bytecode
availability, and call-graph availability. Runtime is min/avg/max across
repositories, in seconds.}
  \label{tab:rq1-results}
  \footnotesize
  \setlength{\tabcolsep}{4pt}
  \begin{tabular}{@{}lrrrrr>{\centering\arraybackslash}p{0.18\columnwidth}@{}}
  \toprule
  \textbf{Build} & \textbf{Repos} & \textbf{B/BC/CG} &
  \textbf{Meth.} & \textbf{Rows} & \textbf{Edges} & \textbf{min/avg/max runtime} \\
  \midrule
  Maven  & \EvalReposMaven{} &
  \EvalMavenBuildSuccess{}/\EvalMavenBytecodeAvailable{}/\EvalMavenCallGraphAvailable{} &
  \EvalMavenMethods{} & \EvalMavenRows{} & \EvalMavenCallEdges{} &
  \EvalMavenRuntimeMin{}/\EvalMavenRuntimeAvg{}/\EvalMavenRuntimeMax{} s\\

  Gradle & \EvalReposGradle{} &
  \EvalGradleBuildSuccess{}/\EvalGradleBytecodeAvailable{}/\EvalGradleCallGraphAvailable{} &
  \EvalGradleMethods{} & \EvalGradleRows{} & \EvalGradleCallEdges{} &
  \EvalGradleRuntimeMin{}/\EvalGradleRuntimeAvg{}/\EvalGradleRuntimeMax{} s\\

  Total  & \EvalReposTotal{} &
  \EvalTotalBuildSuccess{}/\EvalTotalBytecodeAvailable{}/\EvalTotalCallGraphAvailable{} &
  \EvalTotalMethods{} & \EvalTotalRows{} & \EvalTotalCallEdges{} &
  \EvalTotalRuntimeMin{}/\EvalTotalRuntimeAvg{}/\EvalTotalRuntimeMax{} s\\
  \bottomrule
  \end{tabular}
  \end{table}

As shown in Table~\ref{tab:rq1-results}, \tool{} completed the extraction
pipeline for all \EvalReposTotal{} subjects: every repository built, exposed
project bytecode, produced a call graph, and emitted parseable method-context
\jsonl{}. In total, \tool{} emitted \EvalTotalRows{} method-context records and
\EvalTotalCallEdges{} serialized caller/callee entries.

\subsection{RQ2: Source--Bytecode Reconciliation}

  \begin{table}[h]
  \centering
 \caption{RQ2 results. Every serialized caller/callee entry records a bytecode
\texttt{target\_uri}. Project join counts entries linked to a source
\texttt{method\_uri}; project abstention counts recognized project targets
without a unique source match; non-source counts JDK, dependency, and
compiler-generated targets.}
  \label{tab:rq2-results}
  \footnotesize
  \setlength{\tabcolsep}{4pt}
  \begin{tabular}{@{}lrrrrr@{}}
  \toprule
  \textbf{Build} & \textbf{Target URI} & \textbf{Project join} &
  \textbf{Project abst.} & \textbf{Non-source} & \textbf{Recon.} \\
  \midrule

  Maven  & \EvalMavenTargetUri{} & \EvalMavenMethodUri{} &
  \EvalMavenProjectAbstentions{} & \EvalMavenNonProjectEdges{} &
  \EvalMavenProjectSourceRate{} \\

  Gradle & \EvalGradleTargetUri{} & \EvalGradleMethodUri{} &
  \EvalGradleProjectAbstentions{} & \EvalGradleNonProjectEdges{} &
  \EvalGradleProjectSourceRate{} \\

  Total  & \EvalTotalTargetUri{} & \EvalTotalMethodUri{} &
  \EvalTotalProjectAbstentions{} & \EvalTotalNonProjectEdges{} &
  \EvalTotalProjectSourceRate{} \\

  \bottomrule
  \end{tabular}
  \end{table}

Every serialized caller/callee entry preserves the bytecode identity of the
neighboring method as \texttt{target\_uri}. When that target belongs to project
source and has a unique source match, \tool{} adds a source-level
\texttt{method\_uri}. When no unique source declaration is found, \tool{}
abstains and preserves only the bytecode target. Non-source entries include JDK
calls, dependency calls, synthetic bridge methods, and \texttt{invokedynamic}
lambda artifacts.

Table~\ref{tab:rq2-results} shows that all \EvalTotalTargetUri{} entries
preserve a \texttt{target\_uri}. Among
\EvalTotalRecognizedProjectTargets{} recognized project targets, \tool{} joins
\EvalTotalMethodUri{} to a source \texttt{method\_uri} and abstains on
\EvalTotalProjectAbstentions{}, yielding a \EvalTotalProjectSourceRate{}
reconciliation rate. Thus, \tool{} attaches source identities to most recognized
project targets while avoiding unsupported source links.

  \subsection{RQ3: Output Quality}

  \begin{table}[h]
  \centering
  \caption{RQ3 manual audit sample by repository. SLOC counts production source
  lines under \texttt{src/main/java}.}
  \label{tab:rq3-results}
  \footnotesize
  \setlength{\tabcolsep}{5pt}
  \begin{tabular}{@{}lrrrr@{}}
  \toprule
  \textbf{Repo} & \textbf{SLOC} & \textbf{Records} & \textbf{Pass} & \textbf{Rate} \\
  \midrule
  \texttt{MultiPaper} & 6,206 & 5 & 5 & 100.0\% \\
  \texttt{JsonPath} & 10,305 & 7 & 7 & 100.0\% \\
  \texttt{moco} & 15,157 & 11 & 11 & 100.0\% \\
  \texttt{RichTextFX} & 12,076 & 11 & 11 & 100.0\% \\
  \texttt{mockito} & 22,580 & 15 & 15 & 100.0\% \\
  \texttt{jsoup} & 17,933 & 12 & 12 & 100.0\% \\
  \texttt{jedis} & 52,412 & 60 & 59 & 98.3\% \\
  \texttt{spring-cloud-gateway} & 27,146 & 16 & 16 & 100.0\% \\
  \texttt{graphhopper} & 63,322 & 15 & 15 & 100.0\% \\
  \texttt{fastjson2} & 179,175 & 48 & 47 & 97.9\% \\
  \midrule
  \textbf{Total} & 406,312 & 200 & 198 & 99.0\% \\
  \bottomrule
  \end{tabular}
  \end{table}

  To assess output quality, we manually audited \EvalAuditRecords{} method-context
  records sampled from \EvalAuditRepos{} real-world repositories. The audited
  systems span small, medium, and large Java projects, from 6,206 to 179,175
  production SLOC and 406,312 SLOC in total.

  Two independent annotators inspected the same sampled records. A record passed
  only when all applicable checks passed, including method identity,
  documentation references, caller/callee context, and inherited-documentation
  metadata. The annotators agreed on all records, yielding 100.0\% agreement and
  Cohen's $\kappa=1.00$. After adjudication, 198 of 200 records passed, for a 99.0\% overall pass rate.
  Method identity, caller/callee context, and inherited-documentation metadata
  passed for every audited record.

\section{Future Work}

Future work will evaluate whether \tool{} records improve downstream tasks such
as code explanation, documentation generation, test generation, code review, bug
localization, and repair. We also plan to extend the analysis to richer handling
of compiler-generated bytecode artifacts, generated sources, project-specific
layouts, and external dependency documentation. These extensions will preserve
\tool{}'s design principle: attach source identity only when it can be resolved
deterministically; otherwise, retain explicit bytecode identity and abstention
metadata.

\section{Conclusion}

We presented \tool{}, a Java method-context extraction tool for reproducible
code-context mining and dataset generation. \tool{} combines source analysis,
bytecode call graphs, and stable method identities to emit versioned \jsonl{}
records containing source, documentation, structural, metadata, and
caller/callee context. In our \EvalReposTotal{}-repository evaluation, \tool{}
produced \EvalTotalRows{} method-context records and \EvalTotalCallEdges{}
serialized caller/callee entries across Maven and Gradle projects. Every entry
preserved a bytecode \texttt{target\_uri}; among recognized project-source
targets, \tool{} reconciled \EvalTotalMethodUri{} and abstained on
\EvalTotalProjectAbstentions{}, for a \EvalTotalProjectSourceRate{}
reconciliation rate. A \EvalAuditRecords{}-record manual audit found that 99.0\% of the records
  passed the applicable output-quality checks.

We believe, these results position \tool{} as a shared, reproducible foundation for extracting method-level context for LLM-based task automation and for evaluating progress in context-aware software engineering.

\section{Data Availability}
\tool{} is Apache~2.0 open source; code and documentation are available at~\cite{cocomut}, the demonstration video is available at~\cite{tool_demo}, and the archived tool version at the time of submission is available at~\cite{cocomut_zenodo_2026}.

\balance
\bibliographystyle{ACM-Reference-Format}
\bibliography{main}

\end{document}